# A unified theory for exact stochastic modelling of univariate and multivariate processes with continuous, mixed type, or discrete marginal distributions and any correlation structure

Simon Michael Papalexiou

Department of Civil and Environmental Engineering, University of California, Irvine, CA, USA (simon@uci.edu)

**Abstract**
Hydroclimatic processes are characterized by heterogeneous spatiotemporal correlation structures and marginal distributions that can be continuous, mixed-type, discrete or even binary. Simulating exactly such processes can greatly improve hydrological analysis and design. Yet this challenging task is accomplished often by *ad hoc* and approximate methodologies that are devised for specific variables and purposes. In this study, a single framework is proposed allowing the exact simulation of processes with any marginal and any correlation structure. We unify, extent, and improve of a general-purpose modelling strategy based on the assumption that any process can emerge by transforming a "parent" Gaussian process with a specific correlation structure. A novel mathematical representation of the parent-Gaussian scheme provides a consistent and fully general description that supersedes previous specific parameterizations, resulting in a simple, fast and efficient simulation procedure for every spatiotemporal process. In particular, introducing a simple but flexible procedure we obtain a parametric expression of the correlation transformation function, allowing to assess the correlation structure of the parent-Gaussian process that yields the prescribed correlation of the target process after marginal back transformation. The same framework is also applicable for cyclostationary and multivariate modelling. The simulation of a variety of hydroclimatic variables with very different correlation structures and marginals, such as precipitation, stream flow, wind speed, humidity, extreme events per year, etc., as well as a multivariate application, highlights the flexibility, advantages, and complete generality of the proposed methodology.

## 1. Introduction

"*Make things as simple as possible, but not simpler.*" ~Albert Einstein

The stochastic structure and the marginal probability laws of hydroclimatic processes like precipitation, river discharge, temperature, wind, or relevant processes with discrete or binary marginal distributions like the number of extremes per year, wet-dry sequences, etc., show complexities, which the current methodologies are unable to unify in a single modelling framework. These processes differ in their spatiotemporal correlation structures (CS) with more intense structures being observed for example in temperature than in precipitation [*Wilks and Wilby*, 1999]; moreover, they have different marginal distributions, e.g., temperature is normally distributed [*Kilsby et al.*, 2007; *Breinl et al.*, 2015] while the intermittent and highly skewed character of precipitation at fine spatiotemporal scales demands the use of mixed-type distributions [*Stern and Coe*, 1984; *Wilks*, 1998; *Herr and Krzysztofowicz*, 2005; *Papalexiou and Koutsoyiannis*, 2016]. Complexities of these processes are further enhanced by seasonality and diurnal variation, as well as a tendency to normality as we shift from fine to coarser scales due to the central limit theorem. These reasons led to a very large number of stochastic modelling schemes [e.g., *Buishand*, 1978; *Waymire and Gupta*, 1981; *Bardossy and Plate*, 1992; *Onof and Wheater*, 1993; *Wilks*, 1999a; *Onof et al.*, 2000; *Wheater et al.*, 2005; *Grimaldi and Serinaldi*, 2006; *Salas et al.*, 2006; *Cowpertwait et al.*, 2007; *Sirangelo et al.*, 2007; *Bárdossy and Pegram*, 2009; *Mehrotra and Sharma*, 2009; *Serinaldi*, 2009; *Aghakouchak et al.*, 2010; *Lee and Salas*, 2011; *Kleiber et al.*, 2012; *Lombardo et al.*, 2013; *Serinaldi and Kilsby*, 2014b; *Villarini et al.*, 2014; *Hering et al.*, 2015; *Serinaldi and Kilsby*, 2016; *Youngman and Stephenson*, 2016] to name just a few.

Referring to the following sections for technical details, a general unified approach is however possible by assuming that an arbitrary (target) process $\{X(t)\}$ with prescribed marginal distribution $F_X(x)$ and autocorrelation structure (ACS) $\rho_X(\tau)$ has a parent (or else an equivalent) Gaussian process (pGp) $\{Z(t)\}$ with standard Gaussian marginal $\Phi_Z(z)$ and a suitable ACS $\rho_Z(\tau)$. Therefore, modelling and simulating $\{X(t)\}$ only require the definition of two functions $g$ and $\mathcal{T}$ such that $X(t) = g(Z(t))$ and $\rho_Z(\tau) = \mathcal{T}(\rho_X(\tau))$. While $g$ is easily identified as $g(Z) := F_X^{-1}(\Phi_Z(Z))$, this transformation of the marginal distribution does not preserve the ACS but only rank correlation as $g$ is nonlinear and the ACS depends on the marginal distribution $F_X(x)$ [*Embrechts et al.*, 2002]. In particular, for bivariate normally distributed vectors $(Z_i, Z_j)$ and an arbitrary transformation $g: \mathbb{R} \to \mathbb{R}$, one can show that $\rho_X = \text{Cor}[g(Z_i), g(Z_j)] \leq \text{Cor}[Z_i, Z_j] = \rho_Z$ [*Kendall and Stuart*, 1979, p.600]. This means that we need inflated $\rho_Z$ values to obtain the target $\rho_X$. Therefore, the modelling problem reduces to the definition of a correlation transformation function (CTF) $\mathcal{T}$ to estimate the parent-Gaussian autocorrelation (pGACS) $\rho_Z(\tau)$ from a given $\rho_X(\tau)$.

This modeling framework was previously considered in several fields to solve specific problems, thus resulting in quite a sparse literature dealing with simulation of cross-correlated but serially independent realizations from multivariate distributions with specified binary, discrete, and/or continuous marginal distributions [*Emrich and Piedmonte*, 1991; *Macke et al.*, 2009; *Demirtas*, 2014, 2017], or independent time series with specified serial correlation binary, discrete, and/or continuous marginal distributions [*Cario and Nelson*, 1997, 1998; *Kugiumtzis*, 2002; *Macke et al.*, 2009; *Serinaldi and Lombardo*, 2017b, 2017a]. Moreover, these studies focused on specific classes of marginal distributions and CS along with ad hoc solutions for the definition of the CTF $\mathcal{T}$.

In this study, we present a unified and fully general mathematical representation of the parent-Gaussian modeling framework. Going beyond specific cases, we highlight its generality and its ability to model and simulate processes with both cross- and autocorrelations such as spatio-temporal hydrological processes. We introduce a general and efficient procedure to define the function $\mathcal{T}$ based on a simple and flexible parametric function, as well as a general-purpose technique to simulate the pGp with suitably inflated prescribed CS. Finally, we illustrate the performance of the generalized parent-Gaussian framework by using a variety of CS's and marginal distributions of discrete, continuous and mixed type that provide a suitable representation for most of hydroclimatic processes. These heterogeneous examples emphasize how the generalized parent-Gaussian theory can overcome all ad hoc solutions previously proposed in the literature, thus resulting in a powerful but simple tool enabling substantial improvement of hydrological analyses.

The paper is organized as follows: Section 2 describes the theoretical framework introducing a unified mathematical representation of the parent-Gaussian framework starting from the univariate case (i.e., serially correlated processes) and then discussing the multivariate extension. Section 3 starts with a review of marginal distributions and proposes a general framework on how to define and use auto- or cross-correlation structures suitable to describe a large variety of hydroclimatic processes covering most of the cases of practical interest in hydrological studies. Then we present a novel general-purpose CTF along with a general technique to simulate the parent Gaussian process, concluding with the synoptic of the simulation algorithm. Section 4 reports several case studies involving hydroclimatic processes with heterogeneous marginal distributions (discrete, continuous, and mixed) and different CS's, e.g., precipitation, river discharge, wind speed, relative humidity, number of extreme events per year, etc., as well as, a multivariate case of rainfall, wind and relative humidity. Finally, conclusions are summarized in Section 5.

## 2. Theoretical framework

### 2.1 Univariate case

Let $\{X(t) \mid t \in T\}$ be a stationary stochastic process over an indexed set $T$, having an arbitrary autocorrelation structure (ACS) $\rho_{X(t)}(\tau)$ for lag $\tau$, and with the random variable (r.v.) $X(t)$ following an arbitrary marginal distribution function $F_{X(t)}(x)$. Stationarity entails no change in the joint probability distribution of any order and for any shift in time, and thus, $\{X(t)\}$ (for brevity $t \in T$ is omitted hereafter) is defined by a single ACS $\rho_X(\tau)$ and a single marginal distribution $F_X(x)$, as $\rho_{X(t)}(\tau) = \rho_X(\tau)$ and $F_{X(t)}(x) = F_X(x)$ for any $t \in T$. Let $\{Z(t) \mid t \in T\}$ be a standard Gaussian process (Gp) with CS $\rho_Z(\tau)$, i.e., any linear combination of $(Z(t_1), \ldots, Z(t_k))$ has a joint standard Normal distribution [see e.g., *Feller*, 1971, p.87] and any $Z(t)$ follows the standard Normal distribution $\mathcal{N}(0,1)$. Gaussian-processes have well-defined properties and their simulation is easy, therefore, the key concept is to identify a parent Gp that can be transformed to the process $\{X(t)\}$. In essence, if we target to a stochastic process $X(t)$ with a given ACS $\rho_X(\tau)$ and a given marginal $F_X(x)$ we need to transform the pGp, which implies, since the marginal of the Gp r.v.'s is known, to assess the parent-Gaussian process ACS (pGACS) $\rho_Z(\tau)$.

Let $g$ be a real function, $X(t) = g(Z(t))$ a one-to-one transformation transforming $\{Z(t)\}$ into $\{X(t)\}$, and $Z(t) = g^{-1}(X(t))$ its inverse, transforming $\{X(t)\}$ into $\{Z(t)\}$. If such a transformation $g(Z(t))$ exists, when applied to $\{Z(t)\}$, essentially transforms the Gaussian r.v.'s $Z(t) \sim \mathcal{N}(0,1)$ into the r.v.'s $X(t) \sim F_X(x)$, and the pGACS $\rho_Z(\tau)$ into the target process ACS $\rho_X(\tau)$; and vice versa, if $g^{-1}(X(t))$ is applied to $\{X(t)\}$. This creates a unique mapping between the two processes, i.e.,

$$\{Z(t) \mid Z(t) \sim \mathcal{N}(0,1), \rho_Z(\tau)\} \underset{Z(t)=g^{-1}(X(t))}{\overset{X(t)=g(Z(t))}{\rightleftarrows}} \{X(t) \mid X(t) \sim F_X(x), \rho_X(\tau)\} \qquad (1)$$

Now, let us define the transformations

$$X = g(Z) \coloneqq Q_X(\Phi_Z(Z)) \qquad (2)$$
$$Z = g^{-1}(X) \coloneqq Q_Z(F_X(X)) \qquad (3)$$

where $\Phi_Z(z) = 1/2 \, \mathrm{erfc}(-z/\sqrt{2})$ and $Q_Z(u) = -\sqrt{2}\mathrm{erfc}^{-1}(2u)$, with $0 \leq u \leq 1$, are, respectively, the distribution and quantile functions of the standard Gaussian variate $Z$, and $Q_X(u)$ the quantile function of $X$. It is well-known that $Q_X(\Phi_Z(Z))$ maps $Z$ to $X$, with $Q_Z(F_X(X))$ being its inverse transformation mapping $X$ to $Z$. Thus, for the r.v.'s $Z(t) \sim \mathcal{N}(0,1)$ and $X(t) \sim F_X(x)$, where $F_X(x)$ is any well-defined distribution function, a pair of transformation exists transforming one to the other and vice versa.

Let $X(t)$ and $X(\tau) := X(t - \tau)$ (this definition is used for notational simplicity) be a pair of $\{X(t)\}$ r.v.'s at an arbitrary time point $t$, lagged by $\tau$, and having a correlation coefficient $\rho_X(\tau)$ given by

$$\rho_X(\tau) = \frac{E\left((X(t) - \mu_{X(t)})(X(\tau) - \mu_{X(\tau)})\right)}{\sigma_{X(t)}\sigma_{X(\tau)}} = \frac{E(X(t)X(\tau)) - \mu_X^2}{\sigma_X^2} \qquad (4)$$

where $\mu_X$ and $\sigma_X^2$ are the mean and the variance of the r.v. $X$ (given that it exists as there are r.v.'s with infinite mean and variance in which case the correlation cannot be defined), and

$$E(X(t)X(\tau)) = \int_{-\infty}^{\infty}\int_{-\infty}^{\infty} x(t)x(\tau)f_{X(t)X(\tau)}(x(t), x(\tau))\,\mathrm{d}x(t)\mathrm{d}x(\tau) \qquad (5)$$

where $f_{X(t)X(\tau)}(x(t), x(\tau))$ is the joint probability density function (pdf) of $X(t)$ and $X(\tau)$ which, in general, does not have a simple parametric form. Yet since $Z(t)$ and $Z(\tau)$ are jointly normally distributed with correlation $\rho_Z(\tau)$ and with bivariate pdf

$$\varphi_{Z(t)Z(\tau)}(z(t), z(\tau)) = \frac{1}{2\pi\sqrt{1 - \rho_Z(\tau)^2}} \exp\left(-\frac{z(t)^2 - 2\rho_Z(\tau)z(t)z(\tau) + z(\tau)^2}{2(1 - \rho_Z(\tau)^2)}\right) \qquad (6)$$

we can use the theory of r.v.'s transformations [see e.g., *Feller*, 1971] and Eq. (3) to estimate directly the joint pdf of $X(t)$ and $X(\tau)$, i.e.,

$$f_{X(t)X(\tau)}(x(t), x(\tau)) = \varphi_{Z(t)Z(\tau)}\left(g^{-1}(x(t)), g^{-1}(x(\tau)); \rho_Z(\tau)\right) \mathbf{J}(x(t), x(\tau)) \qquad (7)$$

where $\mathbf{J}((x(t), (x(\tau))) = \begin{vmatrix} \partial g^{-1}((x(t))/\partial x(t) & 0 \\ 0 & \partial g^{-1}((x(\tau))/\partial x(\tau) \end{vmatrix}$ is the Jacobian of the transformation [see e.g., *Papoulis and Pillai*, 2002, p.244]. Inspection of (7) reveals that $f_{X(t)X(\tau)}(x(t), x(\tau))$ is a function of the $F_X(x)$ parameters and of $\rho_Z(\tau)$. So, if we insert Eq. (7) into Eq. (5), with a known $\rho_Z(\tau)$ value, known $F_X(x)$ parameters, and calculate numerically the integral, then we can estimate the corresponding $\rho_X(\tau)$ from Eq. (4). This creates a one-to-one link between $\rho_Z(\tau)$ and $\rho_X(\tau)$.

The previous approach has not been proposed before in the literature, and although it results in complex expressions of $f_{X(t)X(\tau)}(x(t), x(\tau))$ affecting the numerical integration speed, in some cases (this was found by empirical trials), e.g., when Gamma type marginals are involved, seems to performs faster than the more standard approach described next. Also, we note that the computation of the Jacobian demands existence of partial derivatives and thus for discrete marginal cases this approach may not be applicable. Another approach to estimate the $\rho_X(\tau)$ is by applying the fundamental probability theorem [*Feller*, 1971, p.5] stating that given an r.v. $Y$ with known pdf $f_Y(y)$ and an r.v. $X = h(Y)$, where $h$ is a real function, the expectation of $X$ can be expressed in terms of $f_Y(y)$, i.e., $E(X) = E(h(Y)) = \int_{-\infty}^{\infty} h(y) f_Y(y) \mathrm{d}y$, avoiding thus, the definition formula $E(X) = \int_{-\infty}^{\infty} x f_X(x) \mathrm{d}x$ which requires knowledge of $f_X(x)$. This theorem can be applied in its two-dimensional form in order to evaluate $E(X(t)X(\tau))$ in terms of the bivariate standard Gaussian pdf (Eq. (6)), and by using the transformation in Eq. (3). Thus, the autocovariance transformation integral $\mathcal{C}(\boldsymbol{\theta}_X, \rho_Z(\tau))$ is given by

$$\mathcal{C}(\boldsymbol{\theta}_X, \rho_Z(\tau)) \coloneqq E(X(t)X(\tau)) = E\left(Q_X\left(\Phi_Z(Z(t))\right) Q_X(\Phi_Z(Z(\tau)))\right)$$
$$= \int_{-\infty}^{\infty} \int_{-\infty}^{\infty} Q_X(\Phi_Z(z(t))) Q_X(\Phi_Z(z(\tau))) \varphi_{Z(t)Z(\tau)}(z(t), z(\tau); \rho_Z(\tau)) \, \mathrm{d}z(t)\mathrm{d}z(\tau) \tag{8}$$

where $\boldsymbol{\theta}_X = (\theta_1, \ldots, \theta_n)$ is the parameter vector of the distribution $F_X(x)$. In general, the double integral $\mathcal{C}(\boldsymbol{\theta}_X, \rho_Z(\tau))$ does not have an analytical solution, yet its numerical estimation is trivial.

Now, if we introduce $\mathcal{C}(\boldsymbol{\theta}_X, \rho_Z(\tau))$ into Eq. (4), we get the autocorrelation transformation integral (ACTI) $\mathcal{R}(\boldsymbol{\theta}_X, \rho_Z(\tau))$, i.e.,

$$\rho_X(\tau) = \mathcal{R}(\boldsymbol{\theta}_X, \rho_Z(\tau)) \coloneqq \frac{\mathcal{C}(\boldsymbol{\theta}_X, \rho_Z(\tau)) - \mu_X^2}{\sigma_X^2} \tag{9}$$

Clearly, $\mathcal{R}(\boldsymbol{\theta}_X, \rho_Z(\tau))$ is a function of $\boldsymbol{\theta}_X$ and of $\rho_Z(\tau)$ as $\mu_X$ and $\sigma_X^2$ are functions of $\boldsymbol{\theta}_X$. Similarly to estimating $\rho_X(\tau)$ by Eq. (4) using direct estimation of $f_{X(t)X(\tau)}(x(t), x(\tau))$, this function creates a link between the $\rho_X(\tau)$ and $\rho_Z(\tau)$, and thus, can be applied for any pair of $(Z(t), Z(\tau))$, or else for all lag values $\tau = 1, \ldots, \tau_{\max}$, transforming essentially the Gaussian ACS $\rho_Z(\tau)$ into the ACS $\rho_X(\tau)$. Note, this relationship transforms a known $\rho_Z(\tau)$ value into a $\rho_X(\tau)$ value, while our goal is to transform an observed or a given $\rho_X(\tau)$ value into the corresponding $\rho_Z(\tau)$ value; this is described in detail in Section 3.3.

## 2.2 Multivariate and cyclostationary case

The equations presented previously were set for a single stationary stochastic process, yet Eq. (9) can be applied in a more general way. As natural processes vary in space and time, in many cases, we need a multivariate and cyclostationary model [*Salas*, 1980; *Salas et al.*, 1982; *Bartolini et al.*, 1988; *Shao and Lund*, 2004]. Let us assume $n$ different processes (e.g., a single process observed in $n$ different locations, $n$ different processes in the same location, etc.), and $m$ different seasons (e.g., 12 months, 365 days, etc.), and let us denote with $\{X_{i,j}(t)\}$ the $i$-th process at $j$-th season, with $i = 1, \ldots n$, and $j = 1, \ldots, m$, thus, there are $n \times m$ different processes to consider. Accordingly, for each process we assume a marginal distribution function $F_{X_{i,j}}(x)$, an ACS $\rho_{X_{i,j}}(\tau)$, and for every pair of processes with $i \neq j$ a cross-correlation structure (CCS) $\rho_{X_{i,j}X_{k,l}}(\tau)$, where $k = 1, \ldots n$, and $l = 1, \ldots, m$. In this case, Eq. (8) becomes

$$\mathcal{C}_{X_{i,j}X_{k,l}}\left(\boldsymbol{\theta}_{X_{i,j}}, \boldsymbol{\theta}_{X_{k,l}}, \rho_Z(\tau)\right) :=$$
$$\int_{-\infty}^{\infty} \int_{-\infty}^{\infty} Q_{X_{i,j}}\left(\Phi_{Z_{i,j}}(z(t))\right) Q_{X_{k,l}}\left(\Phi_{Z_{k,l}}(z(\tau))\right) \varphi_{Z_{i,j}(t)Z_{k,l}(\tau)}(z(t), z(\tau); \rho_Z(\tau)) \, \mathrm{d}z(t) \mathrm{d}z(\tau) \quad (10)$$

and thus, the cross-correlation transformation integral (CCTI) becomes

$$\rho_{X_{i,j}X_{k,l}}(\tau) = \mathcal{R}\left(\boldsymbol{\theta}_{X_{i,j}}, \boldsymbol{\theta}_{X_{k,l}}, \rho_Z(\tau)\right) := \frac{\mathcal{C}_{X_{i,j}X_{k,l}}\left(\boldsymbol{\theta}_{X_{i,j}}, \boldsymbol{\theta}_{X_{k,l}}, \rho_Z(\tau)\right) - \mu_{X_{i,j}}\mu_{X_{k,l}}}{\sigma_{X_{i,j}}\sigma_{X_{k,l}}} \quad (11)$$

where $\boldsymbol{\theta}_{X_{i,j}}, \boldsymbol{\theta}_{X_{k,l}}$ are parameter vectors of $F_{X_{i,j}}(x)$ and $F_{X_{k,l}}(x)$ distributions, respectively. Thus, $\mathcal{R}\left(\boldsymbol{\theta}_{X_{i,j}}, \boldsymbol{\theta}_{X_{k,l}}, \rho_Z(\tau)\right)$ can be seen as a function transforming a given lag-$\tau$ correlation $\rho_{Z_{i,j}Z_{k,l}}(\tau)$ in a pair of standard Gp's, into the lag-$\tau$ correlation $\rho_{X_{i,j}X_{k,l}}(\tau)$ between $\{X_{i,j}(t)\}$ and $\{X_{k,l}(\tau)\}$ processes. Thus, the difference, compared with the univariate stationary case, is that here we have a set of $n \times m$ processes $\{X_{i,j}(t)\}$ and we want to assess a set of parent Gp's $\{Z_{i,j}(t)\}$, or else, their pGCCS's $\rho_{Z_{i,j}Z_{k,l}}(\tau)$. Given this, a multivariate and cyclostationary Gp can be transformed to the set of $\{X_{i,j}(t)\}$ processes using the corresponding transformations $X_{i,j}(t) = Q_{X_{i,j}}\left(\Phi_{Z_{i,j}}(z(t))\right)$. As a result, the problem of generating timeseries from a multivariate cyclostationary model comprising processes with different characteristics, simplifies in generating timeseries from a multivariate cyclostationary standard Gp.

## 3. From theory to practice

### 3.1 Marginal distributions of hydroclimatic processes

#### 3.1.1 Continuous

Basic hydroclimatic processes like temperature, wind, precipitation, river discharge, relative humidity, etc., have continuous, or, mixed-type (partly continuous and partly discrete) marginal probability distributions. The number of parametric distributions used in the literature to describe hydroclimatic r.v.'s is very large: air temperatures is commonly described by the Gaussian distribution [e.g., *Klein Tank and Können*, 2003]; the two-parameter Weibull has been the prevailing model for wind speeds [e.g., *Justus et al.*, 1978;

*Seguro and Lambert*, 2000], yet Gamma, Lognormal and Pearson type I distributions have also been used for wind [for a review see *Carta et al.*, 2009]; nonzero daily precipitation has been modelled, by Gamma [e.g., *Buishand*, 1978; *Bruhn et al.*, 1980; *Geng et al.*, 1986], Weibull [e.g., *Swift and Schreuder*, 1981; *Wilson and Toumi*, 2005], Exponential and mixed Exponentials distributions [e.g., *Smith and Schreiber*, 1974; *Todorovic and Woolhiser*, 1975; *Wilks*, 1999b], or by heavier-tailed distributions like the Lognormal [e.g., *Swift and Schreuder*, 1981], Kappa [*Mielke Jr*, 1973; *Mielke Jr and Johnson*, 1973; *Hosking*, 1994; *Park et al.*, 2009], and the generalized Beta distribution [*Mielke Jr and Johnson*, 1974] (for a detailed global investigation on the distribution of seasonal daily precipitation see *Papalexiou and Koutsoyiannis* [2016] where the Generalized Gamma and the Burr type XII have been used; for global analyses on precipitation extremes see [*Papalexiou and Koutsoyiannis*, 2013; *Papalexiou et al.*, 2013; *Serinaldi and Kilsby*, 2014a] ); river flows have been described by the Lognormal and Pareto distributions [e.g., *Bowers et al.*, 2012]; the Pearson Type III and the 3-parameter lognormal distributions were suggested for low stream flows [*Kroll and Vogel*, 2002]; the Beta distribution has been found to describe very well relative humidity values [e.g., *Yao*, 1974], and also, is a candidate model for percentage r.v.'s like probability dry, percentage of area affected by extreme temperature or precipitation, etc.

    A complete review on the probability models applied in geophysical and hydroclimatic processes would be a difficult task by itself and out of the scope this study. A common practice is to fit many distributions and select the best fitted, yet this approach makes the selection difficult and susceptible to sample variations. In an attempt to simplify this procedure, and based on entropy information measures, a general theoretical framework was proposed [*Papalexiou and Koutsoyiannis*, 2012] that led to consistent distributions for positive geophysical and hydroclimatic r.v.'s. Particularly, the Generalized Gamma ($\mathcal{GG}$) [*Stacy*, 1962; *Stacy and Mihram*, 1965] and the Burr type XII distributions ($\mathcal{B}r$XII) [*Burr*, 1942; *Tadikamalla*, 1980] were derived, yet the $\mathcal{B}r$III is also proposed here as a general and flexible power-type model. Their probability distribution functions are given by

$$f_{\mathcal{GG}}(x;\beta,\gamma_1,\gamma_2) = \frac{\gamma_2}{\beta\,\Gamma(\gamma_1/\gamma_2)}\left(\frac{x}{\beta}\right)^{\gamma_1-1}\exp\left(-\left(\frac{x}{\beta}\right)^{\gamma_2}\right) \qquad (12)$$

$$F_{\mathcal{B}r\text{XII}}(x;\beta,\gamma_1,\gamma_2) = 1 - \left(1+\gamma_2\left(\frac{x}{\beta}\right)^{\gamma_1}\right)^{-\frac{1}{\gamma_1\gamma_2}} \qquad (13)$$

$$F_{\mathcal{B}r\text{III}}(x;\beta,\gamma_1,\gamma_2) = \left(1+\frac{1}{\gamma_1}\left(\frac{x}{\beta}\right)^{-\frac{1}{\gamma_2}}\right)^{-\gamma_1\gamma_2} \qquad (14)$$

These distributions, defined for $x \geq 0$, have scale parameter $\beta > 0$, and shape parameters $\gamma_1 > 0$ and $\gamma_2 > 0$ that control the left and right tails, respectively, providing great shape flexibility. For $\gamma_1 < 1$ and the pdf's are J-shaped and for $\gamma_1 > 1$ bell-shaped, while the parameter $\gamma_2$ controls mainly the "heaviness" of the right tail, and thus, the behaviour of extreme events. Note that $\mathcal{GG}$ is of exponential form having all of its moments finite, while $\mathcal{B}r$XII and $\mathcal{B}r$III are power-type distributions, that can have very heavy tails and infinite moments. The $\mathcal{GG}$ can be considered as a generalization of the two-parameter Gamma and Weibull distributions, while the $\mathcal{B}r$XII includes as special cases the Pareto type II for $\gamma_1 = 1$, or the Weibull distribution as limiting case for $\gamma_2 \to \infty$. Clearly, the principle of parsimony, which requires to always seek for the simplest model, i.e., to use the smallest number of parameters possible [see e.g., *Box et al.*, 2008, p.16], should always be applied. For example, if the Weibull distribution performs satisfactorily (under desired criteria) then it should be preferred over the three-parameter $\mathcal{GG}$; the same holds for the power-type distributions $\mathcal{B}r$XII and $\mathcal{B}r$III where the simpler Pareto type II should be the first choice under inadequate performance.

### 3.1.2 Mixed type, discrete and binary

Several natural processes like precipitation at fine temporal scales (e.g., at daily or subdaily scales), discharge of small streams, or even wind when calm is measured, are intermittent processes. Thus, their marginal distribution comprises a probability mass concentrated at zero (e.g., probability dry in precipitation), and a continuous part for the positive values, expressed by a parametric distribution function. The framework to simulate intermittent processes is the same, (see section 2); the difference lies in the expressions of the distribution function $F_X(x)$ and the quantile function $Q_X(u)$ that now are related to the conditional expressions for $X|X > 0$.

Particularly, if $P(X = 0) = p_0$ is probability zero, then the cdf, pdf and quantile function of $X$, are given, respectively, by

$$F_X(x) = (1 - p_0)F_{X|X>0}(x) + p_0 \quad x \geq 0 \tag{15}$$

$$f_X(x) = \begin{cases} p_0 & x = 0 \\ (1 - p_0)f_{X|X>0}(x) & x > 0 \end{cases} \tag{16}$$

$$x_u = Q_X(u) = \begin{cases} 0 & 0 \leq u \leq p_0 \\ Q_{X|X>0}\left(\dfrac{u - p_0}{1 - p_0}\right) & p_0 < u \leq 1 \end{cases} \tag{17}$$

(the Dirac delta notation can also be used to define the pdf). Additionally, the expressions of the mean and variance of $X$, needed to estimate the CTF's in Eq. (9) or Eq. (11) become

$$\mu_X = (1 - p_0)\mu_{X|X>0} \tag{18}$$

$$\sigma_X^2 = (1-p_0)\sigma_{X|X>0}^2 + p_0(1-p_0)\mu_{X|X>0}^2 \tag{19}$$

The probability zero, can be easily assessed from the empirical data as $p_0 = n_0\backslash n$, with $n_0$ and $n$ denoting, respectively, number of dry and total days, while for positive values any parametric continuous distribution $F_{X|X>0}(x)$ can been used, given that describes sufficiently the empirical data. Obviously, this framework, can be extended to include cases where values above a threshold are considered, i.e., the discrete part can express the $\Pr(X < x_p)$ for an arbitrary value $x_p$, and the continuous part values for $X|X > x_p$. Finally, more general mixed-type marginal distributions can be considered, e.g., the water storage in a reservoir varies theoretically from zero to a maximum level when capacity is reached, thus, as a variable it may have two discrete states, i.e., probability zero and probability for maximum capacity; all in between values can be described by a continuous bounded distribution.

Important aspects of various geophysical process can be considered to have a discrete or a binary marginal distribution [see e.g., *Chung and Salas*, 2000; *Molotch et al.*, 2005; *Serinaldi and Lombardo*, 2017b]. For example, the number of consecutive wet and dry spells (in days, hours, etc.), the number of days per year having a specific property, e.g., temperature, precipitation, or wind speed above a threshold (extremes), the number of wet days per year, binary sequences to describe wet and dry days, and many more. As in the mixed-type case the methodology remains the same, modified only by using expressions for the quantile, mean and variance that correspond to the discrete or the binary probability mass function chosen.

### 3.2 Autocorrelation (ACS) and cross-correlation structures (CCS)

Sample autocorrelation (or cross-correlation) $\hat{\rho}(\tau)$, especially for small samples and for large lags $\tau$, is sensitive to sample variations and calculated up to a maximum lag (up to 1/3 of the sample size is usually suggested). A better approach may be to fit an ACS to $\hat{\rho}(\tau)$, expressed by a simple parametric function $\rho(\tau; \boldsymbol{\theta})$, where $\boldsymbol{\theta} = (\theta_1, \ldots, \theta_k)$ is the parameter space, essentially interpolating the values of $\hat{\rho}(\tau)$ and providing a smoother result. Also, it could be desired to assign a prescribed ACS or CCS e.g., in sensitivity analysis scenarios or if regional information is transferred to a location of interest.

As in Waldo Tobler's First Law of Geography, stating that "near things are more related than distant things", it is well-known that near in time "things" are more related than distance in time "things". Thus, we can assume that any monotonically decreasing function of lag $\tau$ (or distance) having $\rho(0; \boldsymbol{\theta}) = 1$ and $\lim_{\tau \to \infty} \rho(\tau; \boldsymbol{\theta}) = 0$ can serve as a proper ACS. A general class of functions, proposed here, with these properties are the complementary cumulative distribution functions (ccdf), but now treated as functions and not as probability laws. Accordingly, the Weibull ACS is then given by

$$\rho_{\mathrm{W}}(\tau; b, c) = \exp\left(-\left(\frac{\tau}{b}\right)^c\right) \tag{20}$$

with $b > 0$ and $c > 0$, and forms a generalization of the celebrated Markovian ACS as it is identical for $c = 1$, yet decays slower for $c < 1$, and faster for $c > 1$. Note that $\rho_{\mathrm{W}}(\tau)$ can be expressed also in terms of lag-1 correlation $\rho_1$, i.e., $\rho_{\mathrm{W}}(\tau) = \rho_1^{\tau^c}$. A similar stretched exponential form was used for space correlation of rainfall [*Ciach and Krajewski*, 2006]. Likewise, a power-type ACS is formed by the Pareto II ccdf, i.e.,

$$\rho_{\mathrm{PII}}(\tau; b, c) = \left(1 + c\frac{\tau}{b}\right)^{-1/c} \tag{21}$$

with $b > 0$ and $c > 0$. The PII ACS decays always slower than the Markovian, yet it becomes identical as $c \to 0$. A similar two-parameter power-type ACS, named Cauchy class, was also proposed by *Gneiting and Schlather* [2004] as generalization of the fractional Gaussian noise (fGn) ACS.

The Weibull and PII ACS's can be merged into one three-parameter expression, which coincides with the Burr type XII ccdf, i.e.,

$$\rho_{\mathrm{BrXII}}(\tau; b, c_1, c_2) = \left(1 + c_2\left(\frac{\tau}{b}\right)^{c_1}\right)^{-\frac{1}{c_1 c_2}} \tag{22}$$

with $b > 0$, $c_1 > 0$ and $c_2 > 0$; for $c_2 \to 0$, $\rho_{\mathrm{BrXII}}(\tau; b, c_1, c_2) = \rho_{\mathrm{W}}(\tau; bc_1^{1/c_1}, c_1)$, and for $c_1 = 1$ it equals the PII ACS. This general expression, with a different parametrization, was also used in *Papalexiou et al.* [2011], yet it was not suggested within the general framework proposed here and was not identified as the Burr type XII ccdf. Note that we can form ACS's decaying even slower than power-type ACS's, e.g., a general logarithmic expression (GL) proposed here, generalizing also the Markovian ACS, is

$$\rho_{\mathrm{GL}}(\tau; b, c) = \left(1 + \ln\left(1 + c\frac{x}{b}\right)\right)^{-1/c} \tag{23}$$

with $b > 0$ and $c > 0$; for $c \to 0$ coincides with the Markovian ACS, i.e., $\lim_{c \to 0} \rho_{\mathrm{GL}}(\tau; b, c) = \exp(-x/b)$. This ASC for large values of $c$ can have extremely slow decay rates, and may be useful for processes at the second or millisecond timescales.

Other noteworthy ACS are the celebrated fGn, or, Hurst ACS [e.g., *Beran*, 1994, p.52], linked to long-term persistence (LTP) and fGn processes [*Hurst*, 1951; *Mandelbrot and Wallis*, 1968, 1969], given by

$$\rho_{\text{fGn}}(\tau; H) = \frac{1}{2}(|\tau - 1|^{2H} - 2|\tau|^{2H} + |\tau + 1|^{2H}) \sim t^{2H-1} \tag{24}$$

where $0 < H < 1$ is the Hurst coefficient. Also, FARIMA models [*Granger and Joyeux*, 1980; *Hosking*, 1981] have a similar ACS, asymptotically equivalent to the $\rho_{\text{fGn}}(\tau)$, yet their ACS is more flexible as it allows to control short-term correlations; for applications in hydrology see e.g., [*Hosking*, 1984; *Montanari et al.*, 1997]. Various other correlation function have been proposed, for example for wind and atmospheric data [e.g., *Buell*, 1972; *Gneiting*, 1999].

In the cross-correlation case between two process, e.g., of $\{X_i(t)\}$ and $\{X_j(t)\}$, since $\rho_{X_iX_j}(\tau) \neq \rho_{X_iX_j}(-\tau)$ for $\tau \neq 0$, and $\rho_{X_iX_j}(0) \leq 1$ we can still use parametric CCS's similar to the ACS's framework, if this is desired, by modifying them as

$$\rho_{X_iX_j}(\tau) = \begin{cases} \rho_{\text{CS}}(\tau + 1; \boldsymbol{\theta}) & \tau \geq 0 \\ \rho_{\text{CS}}(1 - \tau; \boldsymbol{\theta}) & \tau \leq 0 \end{cases} \tag{25}$$

where $\rho_{\text{CS}}(\tau)$ can be any parametric CS like the Weibull, PII, etc. with different parameter values for $\tau \geq 0$ and $\tau \leq 0$ to fulfil the condition of $\rho_{X_iX_j}(\tau) \neq \rho_{X_iX_j}(-\tau)$. Of course an infinity number of ACS or CCS's can be formed, with many more parameters, yet according to the parsimony principle we should choose the one with the less possible parameters [*Box et al.*, 2008, p.16].

### 3.3 Auto- and cross-correlation transformation function

#### 3.3.1 Autocorrelation transformation function (ACTF)

The ACTI in Eq. (9) transforms a specific autocorrelation value $\rho_Z$ between a pair of $Z(t)$ and $Z(\tau)$ r.v.'s of the Gp lagged by $\tau$, into the autocorrelation $\rho_X$ between the $X(t)$ and $X(\tau)$ r.v.'s, of the target process, yet the inverse relationship is desired for all practical purposes. That is an ACTF $\mathcal{T}(\rho_X)$ where we introduce any value of $\rho_X$ or any ACS $\rho_X(\tau)$ and get, respectively, the pGp $\rho_Z$ value or the pGACS $\rho_Z(\tau)$. We assume here positive correlations ranging in [0,1], yet the framework can be expanded easily for negative correlations too. Essentially, we evaluate the ACTI in a small set of $\rho_Z$ values to create $(\rho_X, \rho_Z)$ points and interpolate them by fitting a simple parametric function. Since $\rho_Z(\tau) \geq \rho_X(\tau)$ [*Kendall and Stuart*, 1979, p.600] the ACTF $\mathcal{T}$ should have the potential to be concave and also pass from the (0,0) and (1,1) points, as for $\rho_Z$ equal to 0 and 1, $\rho_X$ is also 0 and 1.

A simple two-parameter function with these properties we propose here, heuristically evaluated and found to perform perfectly in all cases of continuous and mixed-type marginals studied, is

$$\rho_Z = \mathcal{T}(\rho_X; b, c) = \frac{(1 + b\rho_X)^{1-c} - 1}{(1 + b)^{1-c} - 1} \tag{26}$$

which is concave for $b > 0$ and $c \geq 0$, while for $b \to 0$ and $c = 1$ we have $\rho_Z = \rho_X$. Unless stated this will be the basic ACTF used in the examples of this study. The previous function performs also well for the discrete and binary case, yet the alternative

$$\rho_Z = \mathcal{T}(\rho_X; b, c) = 1 - (1 - \rho_X^b)^c \tag{27}$$

seems to perform better for discrete cases and especially for the binary case. This function is concave for $0 < b \leq 1$ and $c \geq 1$; for $b = 1$ and $c = 1$ we have $\rho_Z = \rho_X$. Note that Eq. (27) coincides with Kumaraswamy cdf [*Kumaraswamy*, 1980], yet here is used as parametric function and not as a probability law. Thus, we can now form a one-to-one relationship that enables the calculation of the pGp $\rho_Z$ either as a single value or as a parametric pGACS if a parametric ACS $\rho_X(\tau)$ is plugged into Eq. (26) or Eq. (27). Any of the ACTF's $\mathcal{T}(\rho_X; b, c)$ functions can be easily fitted if we evaluate the ACTI (Eq. (9)) for $\rho_Z = \{0.1, 0.2, 0.3, 0.4, 0.5, 0.6, 0.7, 0.8, 0.9, 0.95\}$ and use the corresponding set of $(\rho_X, \rho_Z)$ points.

Let us assume we target to a process $\{X(t)\}$ having a Weibull $(\mathcal{W}(\beta, \gamma))$ marginal distribution, $F_X(x) = F_\mathcal{W}(x) = 1 - \exp(-(x/\beta)^\gamma)$ and a Markovian ACS $\rho_X(\tau) = \rho_M(\tau; \rho_1) = \rho_1^\tau$ with $\rho_1 = 0.8$. For illustration and comparison we assume three different $\mathcal{W}$ distributions (Fig. 1a) with scale parameter $\beta = 1$ and shape parameters $\gamma = \{2, 0.5, 0.25\}$, respectively. The estimated $(\rho_X, \rho_Z)$ points (gray dots in Fig. 1b) and the fitted $\mathcal{T}(\rho_X; b, c)$ functions (Fig. 1b) show very different transformation profiles, i.e., for the bell-shaped $\mathcal{W}(1,2)$ which resembles the Gaussian distribution, essentially $\rho_Z = \rho_X$ and the pGACS coincides with $\rho_X(\tau)$ (Fig. 1c). As we deviate from normality, e.g., the $\mathcal{W}(1, 0.25)$ has a heavy tail and very high skewness, the ACTF becomes more concave resulting in very different pGACS's (Fig. 1c). Note that a Markovian ACS $\rho_X(\tau)$ is not reproduced by a Markovian pGACS, e.g., for the $\mathcal{W}(1, 0.25)$ case the $\rho_X(1) = 0.80$ corresponds to $\rho_Z(1) = 0.93$ which does not imply that the Markovian ACS $\rho_Z(\tau) = 0.93^\tau$ will reproduce the $\rho_X(\tau) = 0.80^\tau$, it will only preserve the $\rho_X(1)$.

Another example, showing how the concavity of the ACTF function changes regards mixed-type marginal distributions suitable to model intermittent processes like precipitation (or wind and discharge in some cases). Let us target for an intermittent process $\{X(t)\}$ having probability zero $p_0$, a Pareto II $(\mathcal{P}\text{II}(\beta, \gamma))$ marginal distribution $F_X(x) = F_{\mathcal{P}\text{II}}(x) = 1 - (1 + \gamma x/\beta)^{-1/\gamma}$ for nonzero values with $\beta = 1$ and $\gamma = 0.3$, and an fGn ACS (Eq. (24)) $\rho_X(\tau) = \rho_{\text{fGn}}(\tau; H)$ with $H = 0.75$. For comparison we use $p_0 = \{0, 0.9, 0.99\}$, with the unconditional $\mathcal{P}\text{II}$ pdf's shown in Fig. 1d (note that it is common to observe $p_0$ equal to high values as 0.90 and 0.99 for daily and hourly rainfall records, respectively). The $(\rho_X, \rho_Z)$ points and the fitted $\mathcal{T}(\rho_X; b, c)$ functions (Fig. 1e) show how concavity increases as $p_0$ increases resulting in very different pGACS (Fig. 1c), and thus, for large values of $p_0$ a much stronger $\rho_Z(\tau)$ is necessary to preserve the $\rho_X(\tau)$. As in the previous example the pGACS of the fGn ACS is not an fGN ACS with just a different $H$ value.

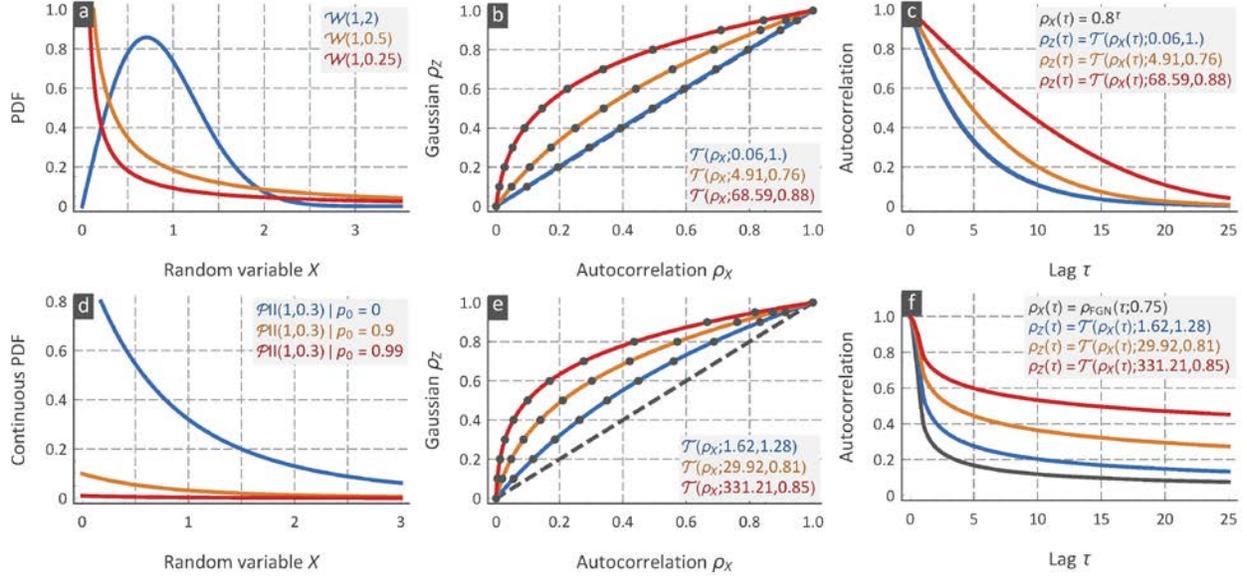

**Fig. 1.** Demonstration of the autocorrelation transformation function (ACTF) and the resulting parent-Gaussian autocorrelation structures (pGACS): (a) different Weibull ($\mathcal{W}$) marginal pdf's, (b) corresponding ACTF's (indicated with the same colour) to the $\mathcal{W}$ distributions, (c) corresponding pGACS assuming $\rho_X(\tau)$ is a Markovian ACS with $\rho_1 = 0.8$ for the three $\mathcal{W}$ distributions (note that $\rho_X(\tau)$ is overlapped by the pGACS of the $\mathcal{W}(1,2)$ case), (d) unconditional Pareto II ($\mathcal{P}$II) pdf's for three $p_0$ values, (e) corresponding ACTF's of the mixed-type $\mathcal{P}$II distributions, (f) corresponding pGACS assuming $\rho_X(\tau)$ is an fGn ACS with Hurst coefficient $H = 0.75$ for the mixed-type $\mathcal{P}$II distributions.

### 3.3.2 Cross-correlation transformation function (CCTF)

In the previous case, the ACTF was constrained to pass form the (1,1) point, yet for the cross-correlation this is not valid. The cross-correlation $\rho_{X_i X_j}$ between two stationary processes $\{X_i(t)\}$ and $\{X_j(t)\}$ has boundaries depending on the marginal distributions $F_{X_i}(x)$ and $F_{X_j}(x)$, thus, there is a range of admissible $\rho_{X_i X_j}$ values according to the Frechet-Hoeffding theoretical limits, i.e., $-1 \leq \rho_{\min} \leq \rho_{X_i X_j} \leq \rho_{\max} \leq 1$ [*Fréchet*, 1957; *Hoeffding*, 1994; *Embrechts et al.*, 2002]. These boundary values can be checked for each specific application by using for example the procedure suggested by *Demirtas and Hedeker* [2011]. However, this check is not required in our approach as the interpolation procedure automatically determines the feasible range of $\rho_{X_i X_j}$ and thus $\rho_{\max}$.

Here, we propose as CCTF a simple two-parameter function, that performed excellent in all cases studied, i.e.,

$$\rho_{Z_iZ_j} = \mathcal{T}\left(\rho_{X_iX_j}; b, c\right) = \left(1 + b\rho_{X_iX_j}\right)^c - 1 \tag{28}$$

Note that if more flexibility is needed the three-parameter $\rho_{Z_iZ_j} = \left(1 + b\rho_{X_iX_j}^{c_1}\right)^{c_2} - 1$ can be used. This function (Eq. (28)) is essentially the same as the ACTF in Eq. (26) but without the denominator part, which forced the function to pass from (1,1); it is valid for $b > 0$ and concave for $c > 1$, linear for $c = 1$, and convex for $c < 1$. As in the ACTF case, we can easily fit the CCTF by evaluating the CCTI in Eq. (11) for $\rho_{Z_iZ_j} = \{0.1, 0.2, 0.3, 0.4, 0.5, 0.6, 0.7, 0.8, 0.9, 0.95, 0.99\}$ and creating the corresponding $(\rho_{X_iX_j}, \rho_{Z_iZ_j})$ points. Once the CCTF parameters are estimated we can determine easily the $\rho_{\max}$ by setting $\rho_{Z_iZ_j} = 1$ in Eq. (28) and solve for $\rho_{X_iX_j}$, i.e.,

$$\rho_{\max} = \frac{2^{1/c} - 1}{b} \tag{29}$$

Let us assume a process $\{X_1(t)\}$ with a Weibull marginal $\mathcal{W}(1, 0.5)$ and a process $\{X_2(t)\}$ with a Beta marginal $\mathcal{B}(\gamma_1, \gamma_2)$. For comparison we use three different Beta pdf's having negative, zero, and positive skewness (Fig. 2a). We use the CCTI of Eq. (11) to estimate the $(\rho_{X_iX_j}, \rho_{Z_iZ_j})$ points (grey dots in Fig. 2b) and we fit the CCTF in Eq. (28). Noteworthy, for the negatively skewed Beta the CCTF is convex, almost a straight line for the symmetric, and concave for the positively skewed Beta, while the corresponding $\rho_{X_iX_j}$ limits, estimated from Eq. (29), are 0.53, 0.64 and 0.75, respectively. This indicates that the more the two marginals differ the more their maximum cross-correlation value $\rho_{X_iX_j}$ will be decreased. We also show the pGCCS of a Markovian CCS (Fig. 2c); clearly, due to the maximum limit not all CCS are feasible.

Another interesting example, is to assume cross-correlated processes with mixed-type marginals that differ in probability zero. We assume a heavy-tailed continuous marginal $\mathcal{P}\mathrm{II}(1, 0.3)$ for the first process $\{X_1(t)\}$, while for the process $\{X_2(t)\}$ we use three different mixed-type $\mathcal{P}\mathrm{II}(1, 0.3)$ marginals with $p_0 = \{0.5, 0.9, 0.99\}$ (Fig. 2d). Using the CCTI of Eq. (11) we calculate the $(\rho_{X_iX_j}, \rho_{Z_iZ_j})$ points (grey dots in Fig. 2e) and we fit the CCTF of Eq. (28). All CCTF's are concave while the corresponding $\rho_{X_iX_j}$ limits, estimated from Eq. (29), are 0.98, 0.89 and 0.66, respectively, indicating a decrease in the maximum cross-correlation as $p_0$ increases, or else, as the two marginal differ more. In Fig. 2f we show the pGCCS of the Markovian CCS used in the previous example; clearly, as $p_0$ increases much stronger pGCCS are needed to reproduce the target CCS.

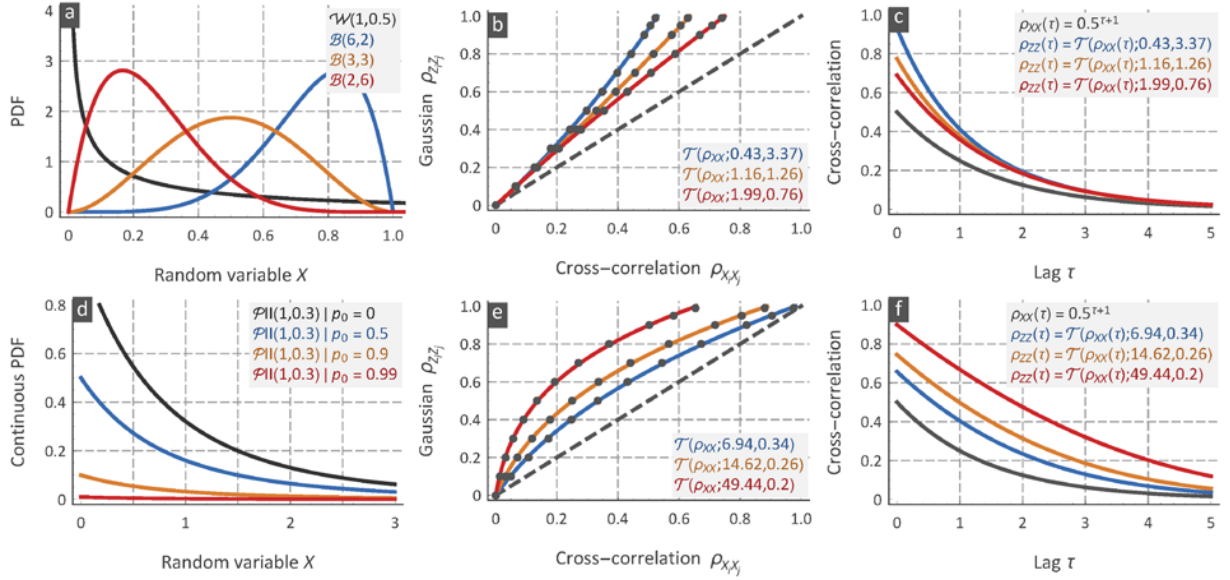

**Fig. 2.** Demonstration of the cross-correlation transformation function (CCTF) and the resulting parent-Gaussian cross-correlation structures (pGCCS): (a) a Weibull ($\mathcal{W}$) marginal and three Beta ($\mathcal{B}$) pdf's, (b) corresponding (indicated with the same colour) cross-correlation transformation functions (CCTF) of the $\mathcal{W}$ distributions with the three Beta, (c) corresponding pGCCS for a Markovian target CCS $\rho_0 = 0.5$, (d) the Pareto II ($\mathcal{P}$II) continuous marginal and the three mixed-type marginal pdf's for three $p_0$ values, (e) corresponding CCTF's, (f) corresponding pGCCS for the Markovian CCS as in the previous example.

### 3.4   Gaussian univariate and multivariate processes with arbitrary CS

We can easily simulate a standard Gp, i.e., with $\mu = 0$ and $\sigma = 1$, having an arbitrary parametric ACS $\rho_Z(\tau)$ by using an autoregressive process of order $p$ (AR($p$)) defined by

$$Z(t) = \sum_{i-1}^{p} a_i Z(t-i) + \varepsilon(t) \tag{30}$$

where $\varepsilon(t) \sim \mathcal{N}(0, \sigma_\varepsilon^2)$ is Gaussian random noise with $\sigma_\varepsilon^2 = 1 - \sum_{i=1}^{p} a_i \rho_Z(\tau)$, and $a_i$ are the model parameters. For a given parametric ACS $\rho_Z(\tau)$ the $a_1, \dots, a_p$ can be easily and analytically estimated using the Yule-Walker equations [e.g., *Box et al.*, 2008, p.67], i.e., $\boldsymbol{a} = \boldsymbol{P}^{-1}\boldsymbol{\rho}$, where $\boldsymbol{a}^{\mathrm{T}} = [a_1, \dots, a_p]$ is the parameter vector, $\boldsymbol{\rho}^{\mathrm{T}} = [\rho_Z(1), \dots, \rho_Z(p)]$ is the correlation vector up to lag $p$, and $\boldsymbol{P} = [\rho_Z(|i-j|)]$ is the $p \times p$ correlation matrix with $i$ and $j$ denoting, respectively, the $i$-th row and $j$-th column, e.g., the $[\boldsymbol{P}]_{2,3}$ element is simply the value of $\rho_Z(1)$ [for more on AR models see *Box et al.*, 2008]

Thus, an AR($p$) model preserves exactly a given ACS $\rho_Z(\tau)$ up to lag $p$ and then for $\tau \geq p + 1$ decays according to $\rho_{\mathrm{AR}(p)}(\tau) = \sum_{i=1}^{p} a_i \rho_Z(\tau - i)$, a relation that can be easily applied recursively. For example, Fig. 3 shows the approximation up to various lags of two parametric ACS, i.e., of a Weibull with $b = 10$ and $c = 0.5$ (decaying thus slower than the Markovian ACS), and a very strong Pareto II ACS with $b = 5$ and $c = 4$. Clearly, the closer the ACS is to the Markovian the better will be approximated by the $\rho_{AR(p)}(\tau)$ for $\tau \geq p + 1$; in any case, a parametric CS can be approximated up to any desired lags, e.g., even an AR(5000) can be easily fitted. We stress that an AR($p$) model of large or very large order, fitted to a parametric ACS, is a very parsimonious model, as all of its parameters are deterministically and analytically derived from the parametric ACS, e.g., an AR(1000) fitted using a two-parameter ACS is essentially a two-parameter model.

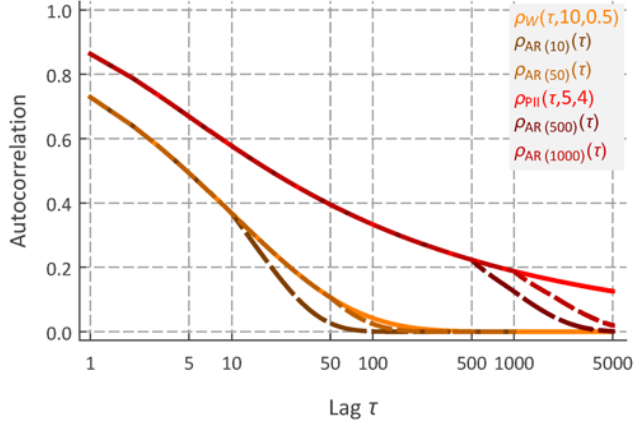

**Fig. 3.** Approximation of two parametric ACS by AR($p$) models of large order.

Another approach to generate timeseries from a standard Gp having an arbitrary ACS $\rho_Z(\tau)$ is based on approximating $\rho_Z(\tau)$ by a sum of independent AR(1) processes, an idea originally proposed and used by *Mandelbrot* [1971] to approximate the fGn ACS. Particularly, if we define

$$Z(t) = \sum_{i=1}^{n} Y_i(t) \tag{31}$$

with $Y_i(t) = \rho_i(1)Y_i(t-1) + \varepsilon_i(t)$ being the $i$-th AR(1) process in the sum, having mean $\mu_{Y_i} = 0$, standard deviation $\sigma_{Y_i}$, lag-1 correlation $\rho_i(1)$, and $\varepsilon_i(t) \sim \mathcal{N}(0, (1 - \rho_i(1)^2)\sigma_{Y_i}^2)$, then, since the AR(1) processes are independent, the variance and the CS of the sum process are $\sigma_Z^2 = \sum_{i=1}^{n} \sigma_{Y_i}^2$ and $\rho_\Sigma(\tau) = 1/\sigma_Z^2 \sum_{i=1}^{n} \rho_i(1)^\tau \sigma_{Y_i}^2$, respectively. Of course, if we demand $\{Z(t)\}$ to have $\sigma_Z^2 = 1$, then $\sum_{i=1}^{n} \sigma_{Y_i}^2$ is also constrained to equal 1. The AR(1) parameters can be estimated by numerically minimizing an error norm between $\rho_\Sigma(\tau)$ and $\rho_Z(\tau)$ up to a desired lag $\tau$. For an application using five independent AR(1) approximating the ACS given Eq. (22) see *Papalexiou et al.* [2011] (for an approximation of the fGn ACS with three AR(1) see [*Koutsoyiannis*, 2002]).

The corresponding multivariate autoregressive model MAR($p$) of $n$ processes is defined by

$$\boldsymbol{Z}(t) = \sum_{i-1}^{p} \boldsymbol{A}_i \boldsymbol{Z}(t-i) + \boldsymbol{B}\boldsymbol{\varepsilon}(t) \qquad (32)$$

where $\boldsymbol{Z}(t) = [Z_1(t), \ldots, Z_n(t)]^\mathrm{T}$, $\boldsymbol{A}_i$ and $\boldsymbol{B}$ are $n \times n$ parameter matrices, and $\boldsymbol{\varepsilon}(t) = [\varepsilon_1(t), \ldots, \varepsilon_n(t)]^\mathrm{T}$ is an $n$-vector of independent standard Gaussian variables. For a general algorithm to fit MAR($p$) models of large order see [*Neumaier and Schneider*, 2001; *Schneider and Neumaier*, 2001; *Schlögl*, 2006]). Clearly, large order MAR models are complicated and often unnecessary, e.g., when Markovian ACS and CCS are observed a MAR(1) is sufficient. The parameters of the MAR(1) model $\boldsymbol{Z}(t) = \boldsymbol{A}\boldsymbol{Z}(t-1) + \boldsymbol{B}\boldsymbol{\varepsilon}(t)$ can be estimated by using the relations $\boldsymbol{A} = \boldsymbol{K}_Z(0)\boldsymbol{K}_Z^{-1}(1)$ and $\boldsymbol{B}\boldsymbol{B}^\mathrm{T} = \boldsymbol{K}_Z(0)\boldsymbol{A}\boldsymbol{K}_Z^\mathrm{T}(1)$ (thus, $\boldsymbol{B}$ can be found using, e.g., Choleski decomposition), where $\boldsymbol{K}_Z(0)$ and $\boldsymbol{K}_Z(1)$ are the correlation matrices of the $n$ Gaussian processes (see also section 4.4 for an application). Of course cyclostationarity can be introduced by using different MAR processes for each season.

### 3.5 A stochastic modelling recipe step-by-step

The proposed stochastic modelling framework is summarized in the following steps:

1. Assess suitable marginal distributions and correlation structures (auto- and cross-correlations for the multivariate case) from the empirical timeseries.
2. Estimate the ($\rho_X, \rho_Z$) points for the proposed $\rho_Z$ values (see section 3.3.1) using the ACTI in Eq. (9) and fit the parametric ACTF of Eq. (26) (or Eq. (27) for binary marginals) to the points; for the multivariate case estimate ($\rho_{Z_iZ_j}, \rho_{X_iX_j}$) points using the CCTI in Eq. (11) and fit the CCTF in Eq. (28).
3. Insert the target ACS $\rho_X(\tau)$ into the fitted ACTF $\mathcal{T}(\rho_X(\tau); b, c)$ to estimate the pGACS $\rho_Z(\tau)$; for the multivariate case use the fitted CCTF $\mathcal{T}\left(\rho_{X_iX_j}(\tau)\right); b, c\right)$ to find the pGCCS $\rho_{Z_iZ_j}(\tau)$.
4. Use the estimated pGACS and the pGCCS to fit an AR($p$) or MAR($p$) model and generate Gaussian univariate or multivariate timeseries.
5. Transform the Gaussian timeseries using the corresponding transformation $X_i(t) = Q_{X_i}\left(\Phi_{Z_i}(z(t))\right)$ based on the fitted distributions from step 1.

## 4. Real world examples

### 4.1 Storm events of fine temporal resolution

The method is applied to simulate fine temporal (10 seconds resolution) rainfall events, a process that has very strong autocorrelation structure and highly skewed and heavy tailed continuous marginal distribution. The original data are seven storm events recorded at the Hydrometeorology Laboratory at the Iowa University [*Georgakakos et al.*, 1994]. This dataset has been used in other studies, e.g., for a statistical and stochastic analysis see *Papalexiou et al.* [2011], for a multifractal analysis see *Cârsteanu and Foufoula-Georgiou* [1996], and for a wavelet analysis see *Kumar and Foufoula-Georgiou* [1997]. We assume that these events are the outcome of a single process [*Papalexiou et al.*, 2011], and thus, the characteristics of the process are better determined if all events are merged into one set despite their large statistical differences (see Fig. 4a). In any case, the aim here in not a detailed analysis of the set, but rather to illustrate the applicability of the method in a highly non-Gaussian process with strong ACS. In addition, other issues like the bias in the estimation of the ACS are not considered [for these detail see *Papalexiou et al.*, 2011].

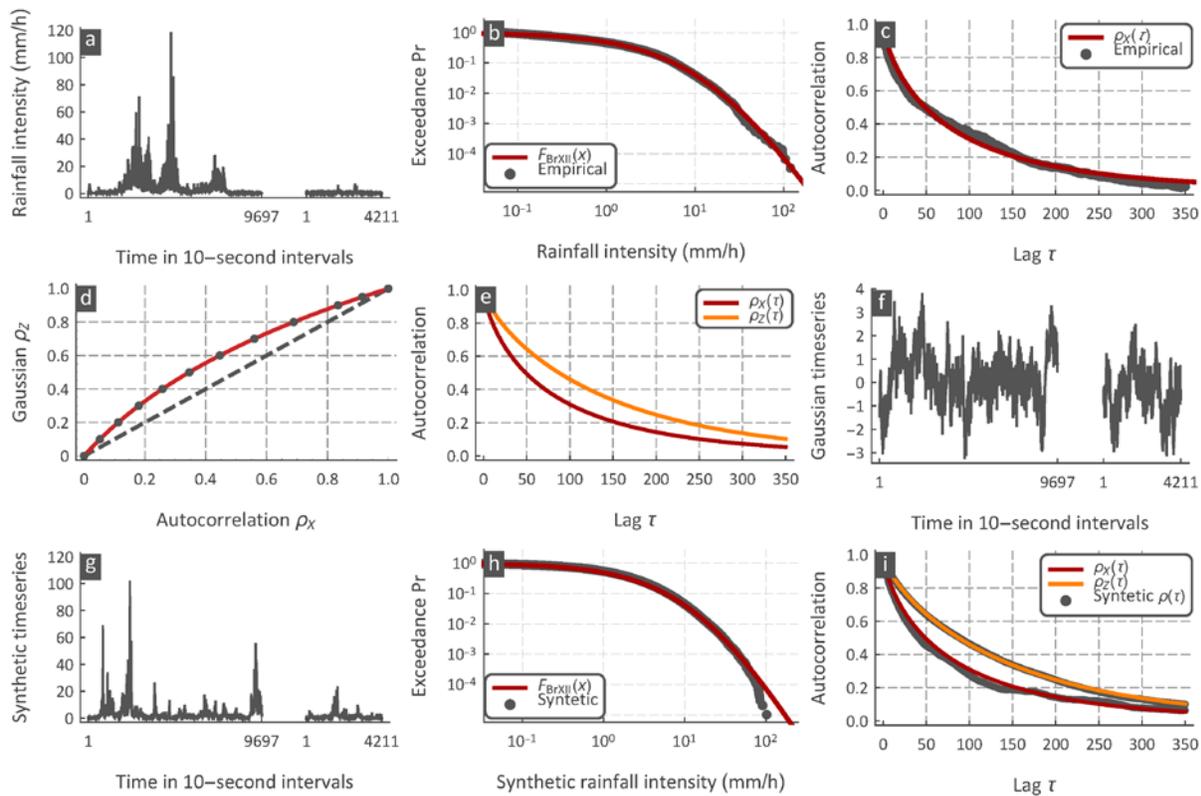

**Fig. 4.** A graphical step by step demonstration of the method applied to 10 s rainfall events: (a) recorded events; (b) empirical distribution of 10 s rainfall and the fitted BrXII distribution; (c) observed ACS and the fitted parametric Weibull ACS $\rho_X(\tau)$; (d) the fitted autocorrelation transformation function (ACTF); (e) the parent-Gaussian autocorrelation structure $\rho_Z(\tau)$; (f) generated Gaussian events with ACS $\rho_Z(\tau)$; (g) synthetic events by transforming the Gaussian timeseries; (h) empirical distribution of the synthetic rainfall compared with the target BrXII fitted to the original data; (i) empirical autocorrelations of Gaussian and synthetic timeseries compared with $\rho_Z(\tau)$ and $\rho_X(\tau)$.

The method is graphically demonstrated in Fig. 4. First, a distribution is fitted to the empirical distribution (Fig. 4b); here the $\mathcal{B}r$XII is used. The estimated parameters, using the method of L-moments [*Hosking*, 1990], are $\beta = 2.13$, $\gamma_1 = 0.74$, and $\gamma_2 = 0.22$ (see *Papalexiou and Koutsoyiannis* [2016] for details in fitting the $\mathcal{B}r$XII using L-moments). Second, the observed ACS $\hat{\rho}_X(\tau)$ is approximated here by numerically fitting the parametric Weibull ACS $\rho_W(\tau)$ given in Eq. (20), with estimated parameters $\beta = 80.6$ and $\gamma = 0.73$ (Fig. 4c), i.e. it is assumed that $\rho_X(\tau) = \rho_W(\tau; \beta, \gamma)$. Third, the ACTI $\mathcal{R}(\boldsymbol{\theta}_X, \rho_Z)$ in Eq. (9) is applied using the fitted $\mathcal{B}r$XII distribution to estimate a few $(\rho_X, \rho_Z)$ points (grey dots in Fig. 4d) which in sequence are "interpolated" by the ACTF in Eq. (26) with parameters $c_1 = 2.96$ and $c_2 = 0.93$. Fourth, the fitted ACTF is applied to the estimated $\rho_W(\tau; \beta, \gamma)$ to obtain the pGACS $\rho_Z(\tau)$ (Fig. 4e). Fifth, a large order AR model is used, i.e., AR(1000), to reproduce the fitted pGACS $\rho_Z(\tau)$ and generate Gaussian TS (Fig. 4f). Sixth, synthetic TS emerge by simply transforming the Gaussian TS using Eq. (2) and the fitted $\mathcal{B}r$XII (Fig. 4e). For verification the empirical distribution of the synthetic timeseries is compared with the target $\mathcal{B}r$XII that was fitted the original data (Fig. 4h), while the empirical ACS of the Gaussian and synthetic timeseries are compared with the theoretical $\rho_Z(\tau)$ and $\rho_X(\tau)$ (Fig. 4i).

### 4.2 Daily precipitation, river discharge and wind

As an example of an intermittent process we use daily rainfall of October recorded at the National Observatory of Athens in Greece from 24/04/1927 to 20/02/1990 (Fig. 5a), assuming that daily rainfall of any month is a stationary process. The probability dry was found $p_0 = 0.78$; the nonzero rainfall was described by an $F_{\mathcal{GG}}(x; 16.5, 0.39, 0.97)$ distribution (Eq. (12)) [*Papalexiou and Koutsoyiannis*, 2016], and a Weibull ACS (Eq.(20)) was fitted to the empirical ACS, i.e., $\rho_X(\tau) = \rho_W(\tau; 0.43, 0.48)$ (dark red line in Fig. 5e). The ACTI in Eq. (9) was used to estimate $(\rho_X, \rho_Z)$ points (grey dots in Fig. 5b) using the corresponding expressions for the mixed-type case, i.e., the quantile function, mean and variance, are given now by Eqs (17), (18), and (19), respectively. The correlation transformation function $\mathcal{T}(\rho_X; 13.88, 0.75)$ fitted perfectly to the $(\rho_X, \rho_Z)$ points (Fig. 5b) and was used to estimate the pGACS, i.e., $\rho_Z(\tau) = \mathcal{T}(\rho_W(\tau; 0.43, 0.48); 13.88, 0.75)$ (orange line in Fig. 5e). The $\rho_Z(\tau) \approx 0$ for $\tau > 20$, thus, an AR(20) was fitted to $\rho_Z(\tau)$ for $\tau = 1, \ldots, 20$ and used to generate a Gaussian TS of 1000 months ($31 \times 1000$ values). The Gaussian TS was transformed to synthetic rainfall (Fig. 5c shows sample size equal with the original TS) by applying the transformation in Eq. (2) with $Q_X$ being the mixed-type quantile in Eq. (17) with the estimated $p_0$ and the fitted $\mathcal{GG}$ quantile. For verification, the empirical distribution and the empirical ACS of the synthetic sample is compared with the expected $F_{\mathcal{GG}}(x; 16.5, 0.39, 0.97)$ distribution (Fig. 5d) and the expected $\rho_X(\tau) = \rho_W(\tau; 0.43, 0.48)$ CS (dark red line in Fig. 5e).

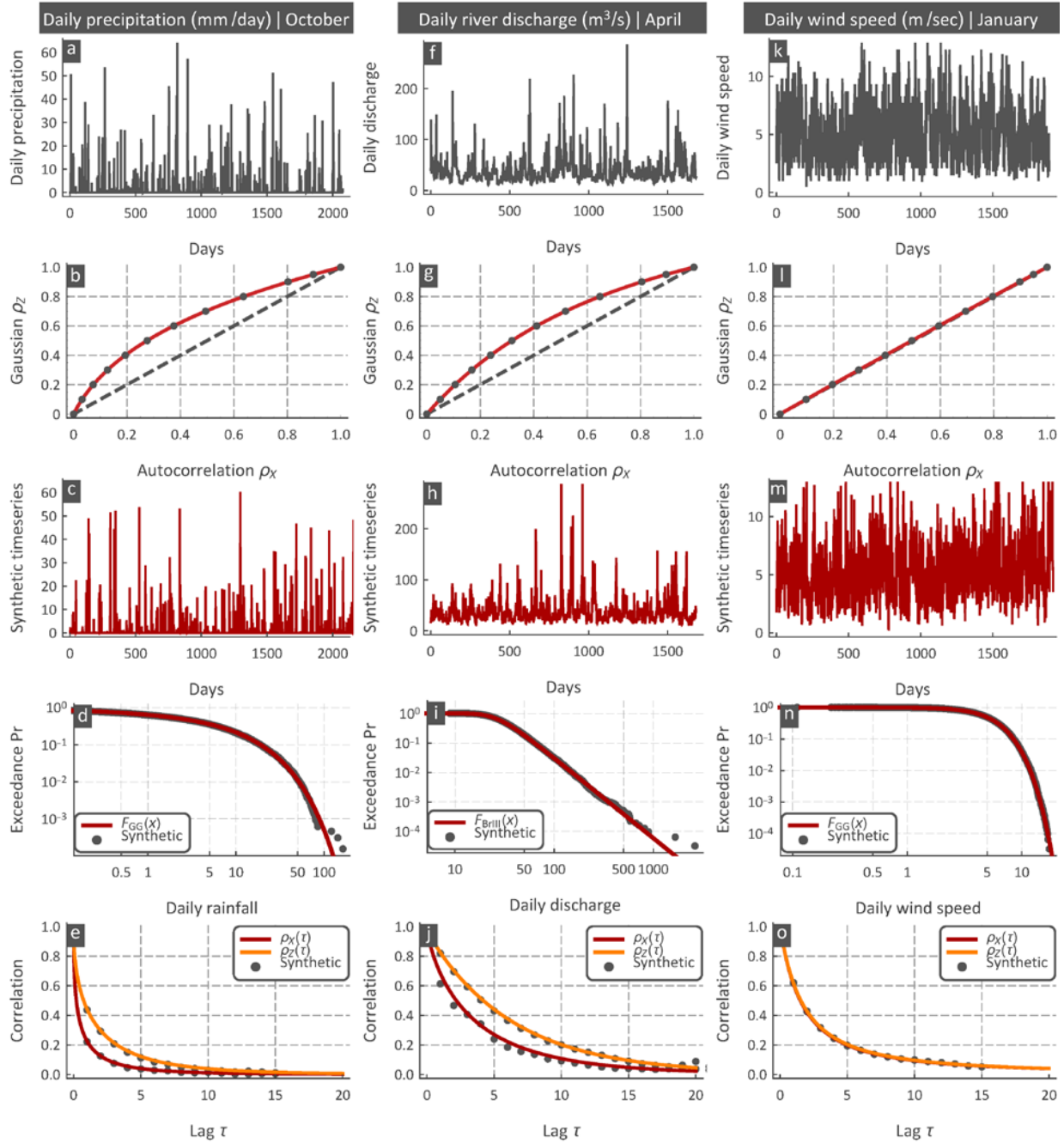

**Fig. 5.** Real-world simulations: (a) daily precipitation of October recorded at the National observatory of Athens, Greece, (b) the corresponding autocorrelation transformation function, (c) simulated timeseries, (d) probability plot verifying that the marginal of the synthetic TS is the same as the fitted $\mathcal{GG}$ to the real nonzero data, (e) the fitted daily precipitation ACS $\rho_X(\tau)$, the estimated pGACS $\rho_Z(\tau)$, and the empirical values from the generated Gaussian and synthetic TS, (f-j) similar to (a-e) for daily river discharge of April in the Middle Fork Snoqualmie river in U.S.A., (k-o) similar to (a-e) for daily wind speed of January recorded at Maastricht, Netherlands.

Another important variable is daily river discharge; here we use daily values of April observed in the Middle Fork Snoqualmie river in U.S.A. from 01/02/1961 to 06/11/2016 (Fig. 5f), assuming again a stationary process for same-month days. Daily discharge was found to be well described by the heavy tailed $F_{\text{BrIII}}(x; 40.5, 12.6, 0.37)$ distribution (Eq.(14)), while the empirical ACS fitted well to a relatively slowly decaying Weibull ACS (Eq.(20)), i.e., $\rho_X(\tau) = \rho_W(\tau; 3.5, 0.79)$ (dark red line in Fig. 5j). The estimated correlation transformation function $\mathcal{T}(\rho_X; 0.74, 2.77)$ (Fig. 5g) was applied to evaluate the pGACS $\rho_Z(\tau)$ (orange line in Fig. 5j), which gets $\rho_Z(\tau) \approx 0$ for $\tau > 25$; thus, a Gaussian TS of 1000 months ($31 \times 1000$ values) was generated from a fitted AR(25) and was transformed to synthetic daily discharge (Fig. 5h) by applying the transformation in Eq. (2) with $Q_X$ being the quantile of the fitted $\mathcal{B}r\text{III}$. The synthetic TS have the expected $F_{\text{BrIII}}(x; 40.5, 12.6, 0.37)$ marginal (Fig. 5i) and the expected $\rho_W(\tau; 3.5, 0.79)$ ACS (dark red line in Fig. 5j).

As a final example of a variable defined in the positive axis the method is applied to daily wind speed of January recorded at Maastricht in the Netherlands from 01/01/1957 to 31/12/2016 (Fig. 5k). Daily wind speed is described here by the $F_{\mathcal{GG}}(x; 4.4, 2.66, 1.76)$ distribution (Eq.(12)), as commonly used models like the two-parameter Weibull distribution or the two-parameter Gamma distribution, did not perform well; also zero values were not recorded and thus the marginal is assumed continuous. The PII ACS in Eq. (21) was fitted to the empirical ACS, i.e., $\rho_X(\tau) = \rho_{\text{PII}}(\tau; 1.7, 0.68)$. The estimated correlation transformation function $\mathcal{T}(\rho_X; 0.31, 0.18)$ (Fig. 5l) shows no concavity, and thus, the pGACS $\rho_Z(\tau)$ coincides with $\rho_X(\tau)$ (lines overlap in Fig. 5o). The $\rho_Z(\tau) \approx \rho_X(\tau) \approx 0$ for $\tau > 25$, and thus, an AR(25) was used to generate a Gaussian TS ($31 \times 1000$ values) which was transformed to synthetic daily wind speed (Fig. 5m) by applying the transformation in Eq. (2) with $Q_X$ being the quantile of the fitted $\mathcal{GG}$. The synthetic TS have the expected marginal (Fig. 5n) and the expected ACS (Fig. 5o).

### 4.3 Relative humidity, discrete and binary cases

Many variables in nature can be expressed in the [0,1] range, e.g., relative humidity, probability dry, percentage variables, etc. Here the method is applied on November daily values of relative humidity recorded at Maastricht in the Netherlands from 01/01/1957 to 31/12/2016 (Fig. 6a). A two-parameter Beta distribution was used, i.e., $F_{\mathcal{B}}(x; 16.1, 2.3)$ and a PII ACS (Eq. (21)) was fitted to the empirical ACS, i.e., $\rho_X(\tau) = \rho_{\text{PII}}(\tau; 0.80, 1, 16)$. The estimated ACTF $\mathcal{T}(\rho_X; 3.24, 0.07)$ (Fig. 6b) is a straight line and the pGACS $\rho_Z(\tau)$ coincides with $\rho_X(\tau)$ (lines overlap in Fig. 6e). The $\rho_Z(\tau) \approx \rho_X(\tau) \approx 0$ for $\tau > 30$, and thus, an AR(30) was used to generate a Gaussian TS ($31 \times 1000$ values) which was transformed to synthetic daily wind speed (Fig. 6c) by applying the transformation in Eq. (2) with $Q_X$ being the quantile of the fitted Beta. The synthetic TS have the expected marginal (Fig. 6d) and the expected ACS (Fig. 6e).

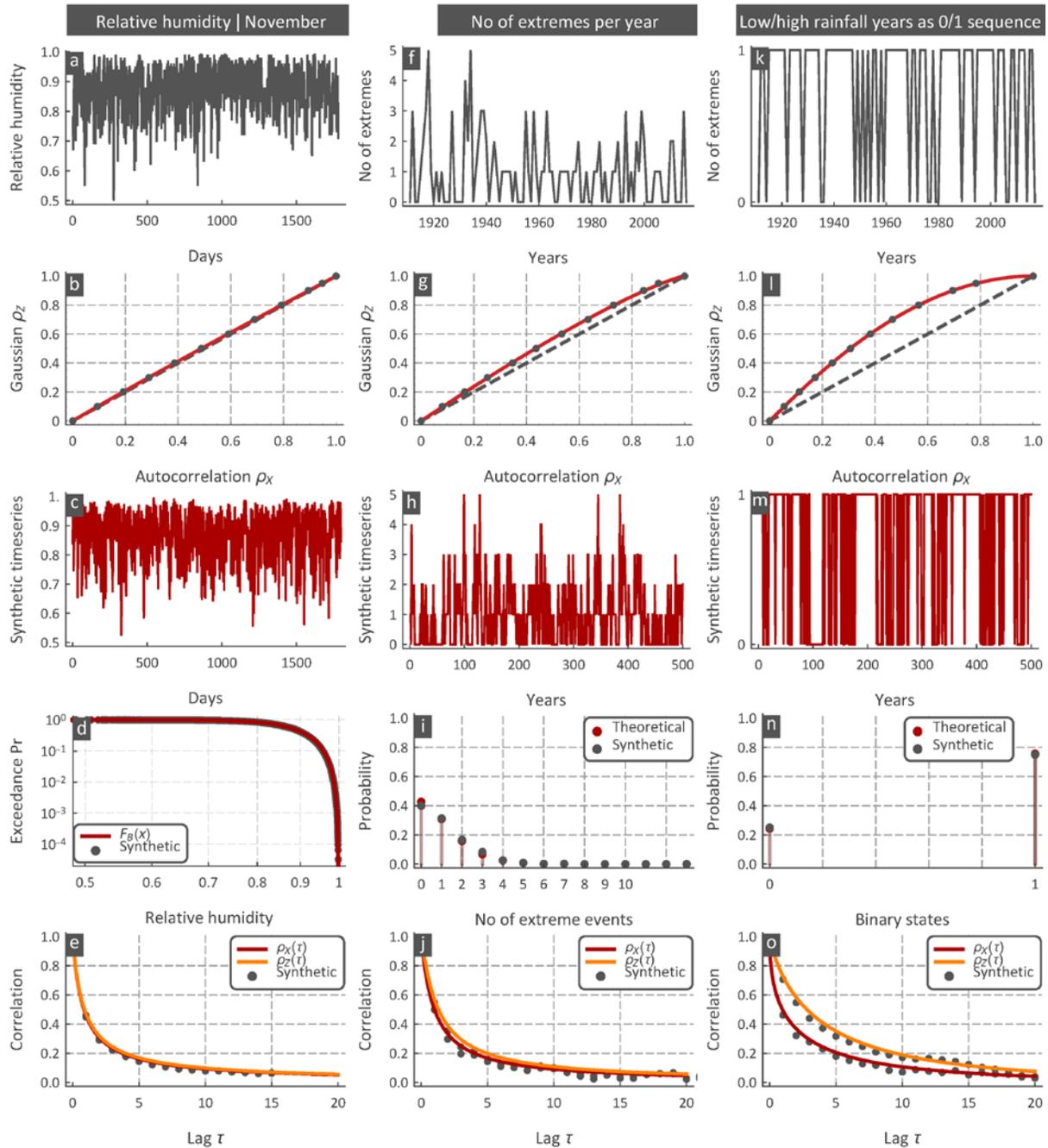

Fig. 6. Simulation of relative humidity, number of extremes per year and occurrences of drought years: (a) relative humidity of November recorded at Maastricht, Netherlands, (b) the ACTF, (c) simulated timeseries, (d) probability plot verifying that the marginal of the synthetic TS is the target Beta distribution, (e) the ACS $\rho_X(\tau)$, the pGACS $\rho_Z(\tau)$, and the empirical values form the generated Gaussian and synthetic TS, (f-j) similar to (a-e) for number of extremes per year in the station ASN00008067 from GHCN-daily database; the discrete Polya-Aeppli distribution is used, (k-o) similar to (a-e) for occurrences of drought years at the station ASN00008067.

As an example of a process with discrete marginal distribution we use the number of extreme daily precipitation events per year, i.e., in an $N$-year record we identify the $N$ largest values and their dates and count the number in each year. Here, we used a long daily rainfall record, randomly selected from the GHCN-Daily database (ID: ASN00008067), to form the number of extremes per year TS (Fig. 6f). A two-parameter discrete distribution was used, i.e., the Polya-Aeppli distribution $F_{\mathcal{PA}}(x; 0.85, 0.15)$, to describe the probability to observe a given number of extremes in any year. The empirical autocorrelation is essentially 0, and thus, for demonstration purposes, a PII CS (Eq. (21)) was prescribed, i.e., $\rho_X(\tau) = \rho_{\text{PII}}(\tau; 1,1)$. Here we fit the Kumaraswamy ACTF in Eq. (27), i.e., $\mathcal{T}(\rho_X; 1.0, 1.22)$ (Fig. 6g) shows small concavity and the pGACS $\rho_Z(\tau) \approx 0$ for $\tau > 30$ (Fig. 6j), and thus, an AR(30) was used to generate a Gaussian TS (3000 years) which was transformed to synthetic number of extremes per year (Fig. 6h) by applying the transformation in Eq. (2) with $Q_X$ being the quantile of the fitted Polya-Aeppli distribution. The synthetic TS have the expected marginal (Fig. 6i) and the expected ACS (Fig. 6j).

In many cases we may desire to express a variable in two states, e.g., rain and no-rain. We use the same daily rainfall record (ASN00008067), aggregated to the annual timescale, to form a binary TS by assigning 0 to annual precipitation values that belong to the first quartile (drought years) and 1 to the rest (Fig. 6k). The Bernoulli distribution was used with $\Pr(X=0) = 0.25$ to describe the probability to observe a drought year, and a Weibull ACS was prescribed, i.e., $\rho_X(\tau) = \rho_W(\tau; 2, 0.5)$ to demonstrate the method as the empirical correlation is almost zero. The estimated Kumaraswamy ACTF $\mathcal{T}(\rho_X; 1.03, 1.97)$ (Fig. 6l) shows moderate to strong concavity with a perfect fit to the points, while the pGACS $\rho_Z(\tau) \approx 0$ for $\tau > 30$ (Fig. 6o). Gaussian TS (3000 years) were generated from an AR(30) and transformed to binary TS accordingly (Fig. 6m). The synthetic binary TS have the expected marginal (Fig. 6n) and the expected ACS (Fig. 6o).

### 4.4 A Multivariate simulation of precipitation, wind, and relative humidity

As a final example we demonstrate the method in a multivariate simulation of three different processes at daily scale, i.e., of precipitation $\{X_1(t)\}$, wind speed $\{X_2(t)\}$ and relative humidity $\{X_3(t)\}$. For precipitation, we assume probability dry $p_0 = 0.7$ and for positive values a power-type $\mathcal{B}r\text{XII}(2, 0.9, 0.2)$ distribution; for wind speed probability calmness $p_0 = 0.1$ and for positive values the two-parameter Weibull $\mathcal{W}(5, 1.2)$ distribution; and for relative humidity a Kumaraswamy $\mathcal{K}u(11, 5)$ marginal distribution (Fig. 7a-c). Thus, we have two mixed-type marginal pdf's, with different $p_0$ and different continuous part, and a continuous marginal bounded in (0,1). Also we assume lag-0 and lag-1 correlation matrices

$$\boldsymbol{K}_X(0) = \begin{pmatrix} 1 & 0.50 & 0.35 \\ 0.50 & 1 & 0.60 \\ 0.35 & 0.60 & 1 \end{pmatrix}, \boldsymbol{K}_X(1) = \begin{pmatrix} 0.30 & 0.25 & 0.15 \\ 0.10 & 0.40 & 0.35 \\ 0.12 & 0.30 & 0.50 \end{pmatrix} \quad (33)$$

with the $(i,j)$ element expressing the cross-correlation coefficients $\rho_{X_iX_j}(\tau)$ for lags 0 or 1 between the $i$-th and $j$-th process. Of course the diagonal in $\boldsymbol{K}_X(1)$ refers to the lag-1 autocorrelation. Note that $\boldsymbol{K}_X(0)$ is always symmetric while $\boldsymbol{K}_X(1)$ is not.

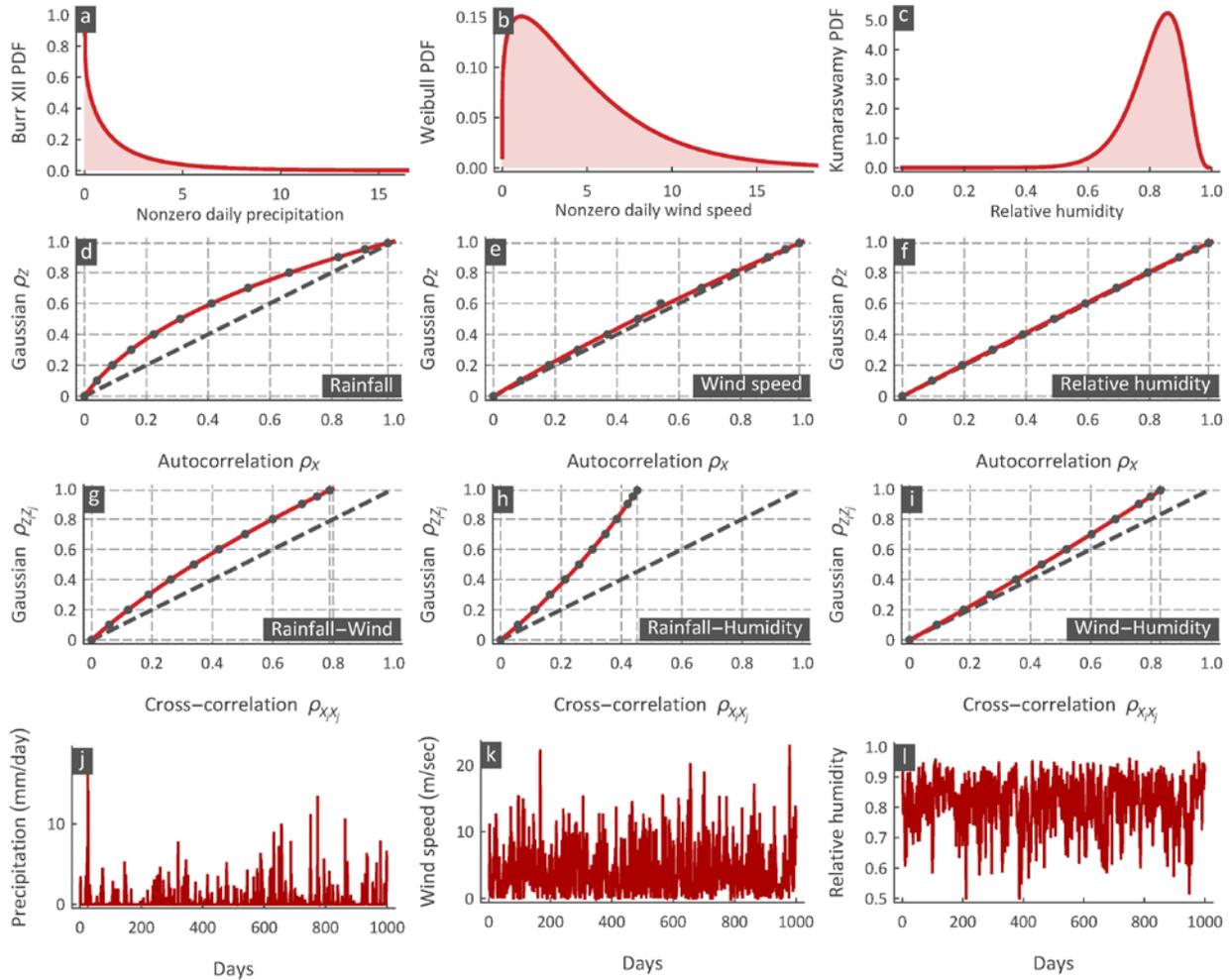

**Fig. 7.** An assumed multivariate case of daily precipitation with $p_0 = 0.7$, wind speed with $p_0 = 0.1$, and relative humidity: (a-c) the fitted or the target marginal pdf's of the three variables, (d-f) the corresponding ACTF's, (g-i) the CCTF's, and (j-l) synthetic times series that preserve the $p_0$, the marginal distributions, the lag-1 autocorrelation, and the lag-0 and lag-1 cross-correlations among them.

The ACTF's (Fig. 7d-f) show a moderate concavity for the precipitation marginal, a slight one for wind, and essentially zero for relative humidity. The CCTF is concave for the rainfall-wind marginal distributions (Fig. 7g) with estimated upper limit 0.80; interestingly, it is clearly convex for the rainfall and humidity with limit 0.42 (Fig. 7h), and slightly convex for the wind-humidity case with limit 0.76 (Fig. 7i). These functions were used to estimate the pGp correlation matrices, i.e.,

$$\boldsymbol{K}_Z(0) = \begin{pmatrix} 1 & 0.69 & 071 \\ 0.69 & 1 & 0.70 \\ 0.71 & 0.70 & 1 \end{pmatrix}, \boldsymbol{K}_Z(1) = \begin{pmatrix} 0.49 & 0.38 & 0.27 \\ 0.17 & 0.44 & 0.40 \\ 0.21 & 0.34 & 0.51 \end{pmatrix} \tag{34}$$

which accordingly were used to fit a multivariate AR(1) and generate a multivariate TS of 10000 values. Accordingly, as in the previous cases, each Gaussian TS was transformed to its target process timeseries using the corresponding transformations $X_i(t) = Q_{X_i}\left(\Phi_{Z_i}(z(t))\right)$. The synthetic TS (Fig. 7j-l) preserve all desired properties, i.e., $p_0$, marginal distributions, lag-1 autocorrelation, and lag-0 and lag-1 cross correlations. Of course, it is straightforward to apply this framework for the multivariate cyclostationary case, and use higher order MAR models if this is desired.

## 5. Summary and Conclusions

Natural processes have a "dynamic" randomness unveiled by structural dependencies in time and/or space and by probability laws governing the frequency of their values. There is no unique correlation structure or probability law to describe the pluralism and the polyphony observed in natural processes. And for this reason, and in order to avoid *ad hoc* solutions, we need a general stochastic framework capable of reproducing this "dynamic" randomness irrespective of the correlation structure or the marginal probability law. The framework presented here, unifies, introduces new elements, extends and simplifies previous attempts and proposes a single approach for stochastic modelling of any hydroclimatic process and beyond.

The basic assumption is that an arbitrary process $\{X(t)\}$, with a prescribed marginal distribution $F_X(x)$ and correlation structure $\rho_X(\tau)$ has a parent-Gaussian process that can be easily assessed. For the multivariate case, this implies that a set of processes have a corresponding set of parent-Gaussian processes. In this direction, a complete framework is introduced based on simple analytical auto- and cross-correlation transformation functions that enable fast and easy estimation of the parent-Gaussian processes avoiding iterative and case specific methods proposed in the past.

The method is applied in a large variety of hydroclimatic processes, ranging from intermittent variables like precipitation and wind speed, to processes with heavy-tail marginals like river discharge, and to processes with bounded marginals like relative humidity, or even to discrete and binary cases. The versatility of the method is also demonstrated in a multivariate case where processes with very different marginals and correlations are reproduced exactly. As a final remark, the framework proposed enables an easy estimation of the limit or the maximum feasible cross-correlation between two processes and reveals that when their marginals differ significantly in shape, they cannot be strongly correlated. The same remark holds also for single processes, and although there is no limit and the autocorrelation can reach up to 1, processes with very different marginals from the Gaussian, e.g., highly skewed and heavy-tailed, will be observed with weak autocorrelation structures, indicating, maybe, that nature likes to "hide" its internal structure.